\documentclass[12pt]{article}
\usepackage{amssymb,amsmath,graphicx}

\begin{document}

\begin{titlepage}

\begin{center}
 
\hfill KEK-TH-1089\\
\hfill \today

\vspace{0.8cm}

{\Large Relic abundance of dark matter
in the minimal universal extra dimension model}

\vspace{1.3cm}

{\bf Mitsuru Kakizaki}$^{a,b,\,}$\footnote{kakizaki@th.physik.uni-bonn.de},
{\bf Shigeki Matsumoto}$^{c,\,}$\footnote{smatsu@post.kek.jp}
and 
{\bf Masato Senami}$^{a,\,}$\footnote{senami@icrr.u-tokyo.ac.jp} \\

\vspace{1cm}

{\it 
$^a${ICRR, University of Tokyo, Kashiwa 277-8582, Japan } \\
$^b${Physikalisches Institut der Universit\"at Bonn,
Nussallee 12, 53115 Bonn, Germany} \\
$^c${Theory Group, KEK, Oho 1-1, Tsukuba, Ibaraki 305-0801, Japan}
}

\vspace{1.5cm}

\abstract{
We investigate the relic abundance of dark matter
in the minimal universal extra dimension model
including resonance processes by second Kaluza-Klein (KK) particles
in all coannihilation processes.
After including second KK resonance processes,
the relic abundance of dark matter is reduced by about 30\%.
Thus, the compactification scale $1/R$ of the extra dimension
consistent with the WMAP observation is increased by a few hundred GeV.
As a result, the cosmologically allowed compactification scale
is $600 {\rm ~GeV} \lesssim 1/R \lesssim 1400 {\rm ~GeV} $ for $\Lambda R = 20$.
}

\end{center}
\end{titlepage}

\section{Introduction}

There is strong evidence of the existence of non-baryonic cold dark matter.
Hence the standard model (SM) must be extended to provide
a dark matter candidate.
Accordingly, many models with a dark matter candidate have been proposed.
Among those, models with a weakly interacting massive particle (WIMP)
are attractive for theoretical reasons.
WIMP can naturally provide the correct relic abundance of
dark matter in the present universe in addition to the successful explanation
of the large scale structure of the universe.

The most extensively studied candidate for WIMP is
the lightest supersymmetric particle (LSP)
in a supersymmetric extension of the SM \cite{susydm},
whose stability is guaranteed by R-parity.
Recently, the lightest Kaluza-Klein particle (LKP)
in the flat universal extra dimension (UED) scenario \cite{Appelquist:2000nn}
has been proposed as an alternative candidate for WIMP.
In UED models, all SM particles can propagate in compact extra dimensions.
The momentum along an extra dimension is interpreted
as the Kaluza-Klein (KK) mass in the four dimensional point of view.
The KK mass spectrum is quantized in terms of KK number $n$ as $m^{(n)} =  n/R$,
where $R$ is the size of the extra dimension.
Hence, the momentum conservation along the extra dimension
is the KK number conservation.
However, the momentum conservation is broken by an orbifold compactification,
which is required for deriving SM chiral fermions as zero modes.
Nevertheless, the KK-parity conservation still remains after the compactification.
Under the parity, the SM particles and even KK particles carry $+1$ charge
while odd KK particles carry $-1$ charge.
As a result, the LKP is stabilized and a good candidate for dark matter.

In this paper we investigate the thermal relic abundance
of the LKP dark matter in the minimal UED (MUED) model.
The MUED model is defined in the five space-time dimensions
in which the extra dimension is compactified on an $S^1/Z_2$ orbifold.
The one-loop corrected KK mass spectrum of the model is calculated
in Ref.~\cite{Cheng:2002iz}.
In the broad parameter region of the model,
the LKP is identified as the KK particle of the photon,
which is dominantly composed of the KK particle of the hypercharge gauge boson,
since the weak mixing angle for the KK particle is negligibly small.

The thermal relic abundance of the LKP produced at the early stage of the universe  
was first investigated in Ref.~\cite{Servant:2002aq}.
In the paper, the relic abundance of the LKP was calculated
including the coannihilation of the KK particles of the SU(2) singlet leptons.
After this work, it is pointed out that
s-channel resonances mediated by second KK particles
are important for the calculation of the relic abundance \cite{Kakizaki:2005en}.
The authors investigated systematically the second KK resonance
for the processes relevant to the LKP and the KK SU(2) singlet leptons
in the paper \cite{Kakizaki:2005uy}.
On the other hand, the relic abundance should be calculated
with coannihilation processes of some other KK particles, 
since their masses are also degenerated with that of the LKP in the MUED model.
This calculation has been done in the papers \cite{Kong:2005hn, Burnell:2005hm}.
However, any second KK resonances has not been included,
and the SM Higgs mass is fixed on 120 GeV. 
Recently, some of us pointed out that the LKP abundance is 
highly dependent on the SM Higgs mass \cite{Matsumoto:2005uh}.
If the SM Higgs mass is larger than 200 GeV,
the KK Higgs coannihilation is efficient 
and the compactification scale $1/R$ consistent with observations
is considerably increased by a maximum of 600 GeV.

The purpose of this paper is investigating
the thermal relic abundance of the LKP in the MUED model
including all second KK resonances in all coannihilation processes
and clarifying dependence of the SM Higgs mass on the abundance.
This paper is organized as follows.
In the next section,
we briefly review the MUED model and its experimental constraint.
In section \ref{Mass spectrum},
we exhibit the mass spectrum of the MUED model, especially
the masses of KK particles of electroweak gauge bosons, Higgs bosons and leptons.
Next we discuss the thermal relic abundance of the LKP dark matter
without second KK particle resonance processes in section \ref{abundance}.
We present the compactification scale consistent with the WMAP observation,
in particular focusing on the dependence of the SM Higgs mass,
and elucidate the importance of the KK Higgs coannihilation.
In section \ref{sec:2ndKK},
we discuss the second KK resonances 
relevant to the calculation of the LKP relic abundance.
Particularly we investigate resonance processes systematically
and clarify which resonance processes are important. 
Then, we calculate the thermal relic abundance of the LKP in the MUED model
including the second KK resonance processes in all coannihilation modes,
and present the allowed region consistent with the observation.
In section \ref{sec:graviton},
we comment on the KK graviton.
Section \ref{summary} is devoted to summary.

\section{Minimal universal extra dimension model}

The simplest version of UED models is called the MUED model.
Hence, this model is adopted as the reference model of UED model by many authors
as the minimal supergravity model in supersymmetric models.
Hence, we investigate the abundance of the LKP in this model,
and extension of the study to other UED models is straightforward.
The MUED model has one extra dimension compactified on an $S^1 / Z_2$ orbifold.
This orbifold compactification is required
for producing SM chiral fermions,
because KK fermions are vector-like four-component fermions.
The orbifold has fixed points,
and the existence of the fixed points in the extra dimension
breaks the translational invariance along the extra dimension.
Hence, the KK number conservation is broken at loop level.
However, the subgroup remains unbroken as KK-parity.
Under this parity,
particles at even (odd) KK levels have plus (minus) charge.
Therefore, the LKP is stable as mentioned in the previous section,
and the first KK particles are always pair-produced in collider experiments.
This situation is quite similar to SUSY models with R-parity,
in which the LSP is stable and supersymmetric particles are pair-produced.

The SM particles and their KK partners have identical charges.
All bulk interactions of the KK particles follow from the SM Lagrangian,
hence, undetermined parameter is only the SM Higgs mass $m_h$.
On the other hand, boundary interactions
cannot be predicted by SM parameters.
Usually, these interactions are fixed at the cut-off scale $ \Lambda $,
which is usually taken to be $\Lambda R \sim {\cal O} (10)$ \cite{Appelquist:2000nn}.
The model is considered to be an effective theory defined at the scale $\Lambda$.
In this paper, we use the boundary interactions adopted in Ref.~\cite{Cheng:2002iz},
which vanish at the scale $\Lambda$
and the effect of them appears through the renormalization group evolution.
As a result, the MUED model is very restrictive and 
has only two new physics parameters,
$\Lambda$ and $1/R$, as well as the SM parameter $m_h$.
Furthermore, our results are almost independent of $\Lambda$,
since it always appears with a loop suppression and
gives only logarithmic correction.
In this paper, we take $\Lambda$ to be $\Lambda R = 20$.

The MUED model is phenomenologically constrained
by $ b \to s \gamma$ \cite{Agashe:2001xt},
the anomalous muon magnetic moment \cite{Agashe:2001ra,Appelquist:2001jz},
$ Z \to b \bar b $ \cite{Appelquist:2000nn,Oliver:2002up},
$B$-$\bar B$ oscillation \cite{Chakraverty:2002qk},
$B $ and $K$ meson decays \cite{Buras:2002ej},
and electroweak precision measurements
\cite{Appelquist:2000nn,Appelquist:2002wb,Flacke:2005hb}.
Among those, the most stringent constraint
comes from the last one in terms of $S$ and $T$ parameters.
The UED contribution to the $ T $ parameter
is known to be canceled somewhat by the SM contribution for heavier Higgs,
and the constraint on $1/R$ depends on the SM Higgs mass 
\cite{Appelquist:2000nn,Appelquist:2002wb}.
For $ m_h \sim 1 $~TeV,
$ 1/R \sim 300 $~GeV is allowed by the electroweak precision measurements.
However, the SM Higgs mass should be less than 300 GeV in the MUED model,
otherwise LKP would be the charged KK Higgs boson as shown in the next section.
Stable charged particle with its mass smaller than 1 TeV
is ruled out by the anomalous heavy water molecule search in sea water \cite{CHAMP}.
The null result requires extremely low abundance,
and charged stable particles are overproduced
if the reheating temperature after the inflation is larger than 1 MeV,
which is inconsistent with the successful big bang nucleosynthesis.
As a result, the electroweak precision measurements constrain
the compactification scale as $1/R \gtrsim 400$~GeV.

\section{Mass spectrum}
\label{Mass spectrum}

The mass spectrum of KK particles in the MUED model has an interesting feature
that all particles are degenerated in mass.
This feature is important for
the calculation of the relic abundance of the LKP,
since the abundance is almost governed by coannihilation processes.

The mass spectrum of KK particles at tree level
are determined by $1/R$ and the masses of corresponding SM particles.
Since $ 1 /R $ is much larger than SM particle masses,
all particles at each KK level are highly degenerated.
In particular, the KK particles of the massless SM particles
are exactly degenerated.
Therefore, we should consider the radiative corrections to their masses
to find out which particle is the LKP.
In the MUED model, the radiative corrections to KK particles
has been studied in Ref.~\cite{Cheng:2002iz}.
We summarize their results relevant to this work.

The LKP turns out to be the first KK photon, $\gamma^{(1)}$,
in a reasonable parameter region in the MUED model.
Since the KK photon is a good candidate for dark matter,
the region is considered to be the most well-motivated.
The masses for the KK photon and the KK $Z$ boson are obtained by diagonalizing
the mass squared matrix described in the $(B^{(n)}, W^{3(n)})$ basis,
\begin{eqnarray}
  \left(
  \begin{array}{cc}
   (n/R)^2 + \delta m^2_{B^{(n)}} + g^{\prime 2} v^2/4 & g'g v^2/4 \\
   g'g v^2/4 & (n/R)^2 + \delta m^2_{W^{(n)}} + g^2 v^2/4
  \end{array}
 \right) ,
 \label{eq:LKP_mass_matrix}
\end{eqnarray}
where $g(g')$ is the SU(2)$_L$(U(1)$_Y$) gauge coupling constant
and $v \simeq 246$ GeV is the vacuum expectation value (vev) of the SM Higgs field. 
The radiative corrections to the KK gauge bosons are given by
\begin{eqnarray}
\delta m^2_{B^{(n)}}
 &=&
 -\frac{39}{2}\frac{g^{\prime 2}\zeta(3)}{16\pi^4 R^2}
 -\frac{g^{\prime 2}}{6} \frac{n^2}{R^2}
  \frac{ \ln \left(\Lambda^2 R^2\right) }{16 \pi^2} ,
\\
\delta m^2_{W^{(n)}} &=&
 -\frac{5}{2}\frac{g^2\zeta(3)}{16\pi^4 R^2}
 +\frac{15 g^2}{2}\frac{n^2}{R^2}
  \frac{ \ln \left(\Lambda^2 R^2\right) }{16 \pi^2} .
\end{eqnarray}
The difference between diagonal elements
exceeds the off-diagonal ones when $1/R \gg v$.
Thus, the weak mixing angle of the KK gauge bosons is tiny,
and the KK photon is dominantly composed of
the KK particle of the hypercharge gauge boson.
The mass of the KK particle of $W^\pm$ boson is given by
\begin{eqnarray}
m_{W^{(n)}}^2 = (n/R)^2 + \delta m^2_{W^{(n)}} + g^2 v^2/4 .
\end{eqnarray}

Next, we discuss masses of the KK Higgs particles.
Since KK modes of the Higgs field are not eaten by SM gauge bosons,
all of them, neutral scalar $H^{(n)}$, pseudoscalar $A^{(n)}$ and
charged scalar $H^{\pm(n)}$, remain as physical states.
The latter two, $A^{(n)}$ and $H^{\pm(n)}$,
are the KK particles of the Goldstone modes in the SM.
For the KK Higgs doublet $ \phi^{(n)} $, we take the following notation,
\begin{eqnarray}
\phi^{(n)} &=& \left( H^{+(n)} , \frac{H^{(n)} + i A^{(n)}}{ \sqrt{2} } \right) .
\end{eqnarray}
The KK Higgs boson masses turn out to be
\begin{eqnarray}
 m_{H^{(n)}}^2 &=& (n/R)^2 + m_h^2 + \delta m_{H^{(n)}}^2 ,
 \label{eq:higgsmass} \\
 m_{H^{\pm (n)}}^2 &=& (n/R)^2 + m_W^2 + \delta m_{H^{(n)}}^2 ,
 \label{eq:chargedhiggsmass} \\
 m_{A^{(n)}}^2 &=& (n/R)^2 + m_Z^2 + \delta m_{H^{(n)}}^2 ,
 \label{eq:pseudohiggsmass}
\end{eqnarray}
where $m_W$ and $m_Z$ are the $W$ and $Z$ boson masses, respectively.
The radiative correction to the KK Higgs bosons is given by
\begin{eqnarray}
 \delta m_{H^{(n)}}^2 = \left( \frac{3}{2}g^2 + \frac{3}{4}g^{\prime 2} - \lambda_h \right)
 \frac{n^2}{R^2} \frac{ \ln \left(\Lambda^2 R^2\right)}{16\pi^2} ,
 \label{eq:deltamh}
\end{eqnarray}
where $\lambda_h$ is the Higgs self-coupling defined as
$\lambda_h \equiv m_h^2/v^2 .$
As increasing $m_h$, $\lambda_h$ becomes large,
and the negative contribution in Eq.~(\ref{eq:deltamh}) increases. 
Hence, for large $\lambda_h$,
the mass difference between the KK photon and $H^{\pm(1)}(A^{(1)})$ becomes small,
while the mass of $H^{(1)}$ becomes large as seen in Eq.~(\ref{eq:higgsmass}).
However, the mass difference is negative when $m_h$ is too large.
The charged KK Higgs LKP is not allowed
from the point of view of dark matter.

\begin{figure}[t]
\begin{center}
\scalebox{.9}{\includegraphics*{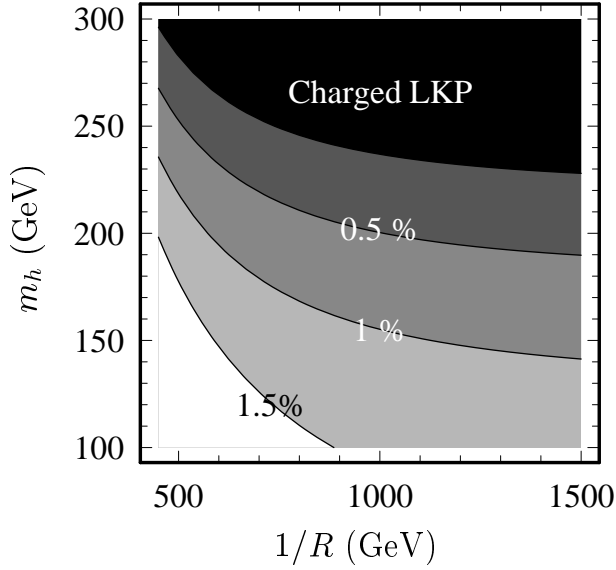}}
\caption{\small The contour map of the mass degeneracy
  between the first KK photon and charged KK Higgs,
  $(m_{H^{(1)\pm}} - m_{\gamma^{(1)}} )/ m_{\gamma^{(1)}}$, in ($1/R$, $m_h$) plane.
  The cut-off scale $\Lambda$ is set to be $\Lambda R =20$.}
\label{fig:lkpcont}
\end{center}
\end{figure}

In Fig.~\ref{fig:lkpcont}, we depict the mass degeneracy
between the LKP and the charged KK Higgs boson, 
$(m_{H^{\pm(1)}} -  m_{\gamma^{(1)}})/ m_{\gamma^{(1)}}$.
It is clear from this figure that
the mass difference between them is very small for $m_h \sim 150  - 230$ GeV.
On the other hand, 
$H^{\pm(1)}$ becomes the LKP for $m_h \gtrsim 230 - 300$ GeV, 
so that this parameter region should be discarded from our discussion.
The bound on the charged LKP region become slightly strong (weak)
by several GeV for larger (smaller) $\Lambda$.

KK leptons are also important for coannihilations,
since they are well degenerated with the LKP
in most of an interesting parameter region.
The masses and eigenstates for KK charged leptons
are obtained from the following mass matrix,
\begin{eqnarray}
  \left(
  \begin{array}{cc}
   n/R + \delta m_{l^{(n)}} & m_l \\
   m_l & - n / R  - \delta m_{L^{(n)}}
  \end{array}
 \right) ,
 \label{eq:KKleptonmass}
\end{eqnarray}
where $l(L)$ represents the SU(2) doublet (singlet) lepton,
and $m_l$ is the corresponding SM lepton mass.
The radiative corrections are given by
\begin{eqnarray}
\delta m_{l^{(n)}} &=&
\left( \frac{27}{16} g^2 + \frac{9}{16} g^{\prime 2} \right)
\frac{n}{R} \frac{ \ln \left(\Lambda^2 R^2\right)}{16\pi^2} , \\
\delta m_{L^{(n)}} &=& \frac{9 g^{\prime 2}}{4} 
\frac{n}{R} \frac{ \ln \left(\Lambda^2 R^2\right)}{16\pi^2} .
\end{eqnarray}
The mass eigenstates of the KK leptons are obtained
by diagonalizing the mass matrix in Eq.~(\ref{eq:KKleptonmass}).
Since the off-diagonal elements are negligibly small compared to the diagonal ones,
the mass eigenstates of KK charged leptons
are given as the SU(2) singlet and doublet leptons.
On the other hand, the KK neutrino masses are given by
\begin{eqnarray}
 m_{\nu^{(n)}} = n/R + \delta m_{l^{(n)}} .
 \label{eq:KKnumass}
\end{eqnarray}
In this work, we do not address the origin of the nonzero neutrino masses.
For example,
even if we include right-handed neutrinos to induce tiny Dirac neutrino mass,
Eq.~(\ref{eq:KKnumass}) remains true,
since the off-diagonal elements are negligibly small as in the charged lepton case.

The others are the KK quarks and gluons.
These are not important for our calculation of the relic abundance of the LKP.
This is because the radiative corrections to their masses
are large due to the strong interaction, and they are heavy so that
coannihilation processes for these particles
are suppressed by Boltzmann factors.
Though we include them in our numerical calculation for completeness,
we only refer Ref.~\cite{Cheng:2002iz} for their masses here.

\begin{table}[tp]
\begin{center}
\begin{tabular}{|c|c|c|}
\hline
& $1/R = 500$GeV  & $1/R = 800$GeV \\
& $ m_h = 120$GeV & $ m_h = 200$GeV  \\
\hline
$\gamma^{(1)}$ & 501 GeV & 800 GeV\\
\hline
$E^{(1)}$ & 506 GeV & 809 GeV\\
\hline
$e^{(1)}$ & 515 GeV & 823 GeV\\
\hline
$\nu^{(1)}$ & 515 GeV & 823 GeV\\
\hline
$H^{\pm(1)}$ & 511 GeV & 805 GeV\\
\hline
$A^{(1)}$ & 513 GeV & 806 GeV\\
\hline
$H^{(1)}$ & 519 GeV & 825 GeV\\
\hline
$W^{(1)}$ & 534 GeV & 849 GeV \\
\hline
$Z^{(1)}$ & 534 GeV & 849 GeV \\
\hline
$U^{(1)}$ & 572 GeV & 910 GeV \\
\hline
$D^{(1)}$ & 571 GeV & 907 GeV \\
\hline
$u^{(1)}$ & 583 GeV & 928 GeV \\
\hline
$d^{(1)}$ & 583 GeV & 928 GeV \\
\hline
$T^{(1)}$ & 571 GeV & 882 GeV \\
\hline
$t^{(1)}$ & 596 GeV & 922 GeV \\
\hline
$g^{(1)}$ & 620 GeV & 983 GeV \\
\hline
$H^{(2)}$ & 1016 GeV & 1614 GeV \\
\hline
$W^{(2)}$ & 1059 GeV & 1691 GeV \\
\hline
$Z^{(2)}$ & 1059 GeV & 1691 GeV \\
\hline
\end{tabular}
\caption{Typical mass spectra are listed.
For KK fermions, small (capital) letters represent SU(2) doublet (singlet) fermions.
We show only the first generation of the KK leptons,
since the masses of the KK leptons are almost independent of their SM mass,
while the masses of the KK top quarks are slightly affected by the SM top mass.
We include some second KK particle masses relevant to our calculation.  }
\label{tab:massspectrum}
\end{center}
\end{table}

In Table~\ref{tab:massspectrum}, the typical mass spectra are shown
for $(1/R,m_h)=$(500 GeV, 120 GeV), (800 GeV, 200 GeV).
In calculation of mass spectra,
we use the gauge couplings improved by the renormalization group running.
For masses of second KK particles,
the first KK particles modify the running of the coupling constants,
thus we used the $\beta$ functions dictated by the first KK particle spectrum
for the running between $1/R$ and $2/R$ \cite{Dienes:1998vg}.

\section{Relic abundance of the LKP dark matter}
\label{abundance}

We are now in a position to calculate
the thermal relic abundance of the LKP dark matter.
Before calculating the relic abundance,
we briefly explain our method to calculate it.
In this section, the discussion is restricted to the case without resonace processes,
which will be included in the next section.
Nevertheless, the result without resonance processes is
a good illustrative example to understand the result
including second KK resonance processes.

\subsection{Boltzmann equation}

We use the method developed in Ref.~\cite{Griest:1990kh,Gondolo:1990dk}
for the calculation of the relic abundance including coannihilation effects.
Under reasonable assumptions,
the relic density obeys the following Boltzmann equation,
\begin{eqnarray}
 \frac{dY}{dx} =
 -\frac{\langle \sigma_{\rm eff} v \rangle}{Hx}
 \left( 1 - \frac{x}{3 g_{*s}} \frac{d g_{*s}}{d x} \right)
 s \left( Y^2 - Y_{\rm eq}^2 \right) ,
 \label{eq:Boltzmann}
\end{eqnarray}
where $\langle \sigma_{\rm eff} v \rangle $ is thermally averaged effective
annihilation cross section defined below, $Y = n/s$ and $x = m_{\gamma^{(1)}} /T$.
The number density $n$ is defined by
the sum of the number density of each species $i$ as $n \equiv \sum_i n_i$.
The abundance in equilibrium is given by
\begin{eqnarray}
 Y_{\rm eq} = \frac{45}{2 \pi^4} \left( \frac{\pi}{8} \right)^{1/2}
 \frac{g_{\rm eff}(x)}{g_{*}(x)} x^{3/2} e^{-x} ,
\end{eqnarray}
where $g_{\rm eff}(x)$ is the number of the effective degrees of freedom of KK particles
and defined by
\begin{eqnarray}
 g_{\rm eff} (x)= \sum_i g_i(1 + \Delta_i)^{3/2} e^{-x \Delta_i} ,
 \qquad
 \Delta_i = (m_i - m_{\gamma^{(1)}})/m_{\gamma^{(1)}} .
 \label{eq:geff}
\end{eqnarray}
The number of the internal degrees of freedom for species $i$ are denoted by $g_i$.
The entropy density is given by
$ s = (2 \pi^2 / 45) g_{*s} (x) m_{\gamma^{(1)}}^3 / x^3 $
and the Hubble parameter is $H = 1.66 g_*(x)^{1/2} m_{\gamma^{(1)}}^2/x^2 m_{\rm Pl}$,
where $m_{\rm Pl} = 1.22\times 10^{19}$ GeV is the Planck mass.
The relativistic degrees of freedom of the thermal bath, $g_*$ and $g_{*s}$,
are treated as the function of temperature in our calculation.
The temperature dependence of $g_{*}$ and $g_{*s}$
is required for deriving the correct abundance.
In the MUED model, $\langle \sigma_{\rm eff} v \rangle $
is dependent on the temperature in a non-trivial way
even though s-channel resonance processes are not included,
hence we solve the Boltzmann equation numerically to sufficiently low temperature.
In section \ref{sec:wo2ndKK}, we discuss this point in detail.

The effective annihilation cross section $\sigma_{\rm eff}$
is given as the sum of $\sigma_{ij}$,
which denotes the coannihilation cross section between species $i$ and $j$,
\begin{eqnarray}
 \sigma_{\rm eff} = \sum_{i,j} \sigma_{ij} \frac{g_i g_j}{g_{\rm eff}^2} 
 (1 + \Delta_i)^{3/2} (1 + \Delta_j)^{3/2} \exp[-x(\Delta_i + \Delta_j)] .
 \label{effective CS}
\end{eqnarray}
All annihilation cross sections, $\sigma_{ij}$,
at tree level has been already calculated.
For the explicit expressions,
see Refs.~\cite{Servant:2002aq,Kong:2005hn,Burnell:2005hm}.
In our work, we include some loop level diagrams,
which is given in section \ref{sec:2ndKK}.
The thermal average for a resonance process
is calculated by the method in Ref.~\cite{Gondolo:1990dk}.

In this paper,
we present the relic density in terms of $\Omega h^2 = m n h^2 / \rho_c$,
which is the ratio of the dark matter density to the critical density
$\rho_c = 1.1 \times 10^{-5}\ h^2\ {\rm GeV cm}^{-3}$. 
The small letter $h$ denotes the scaled Hubble parameter,
$h = 0.73^{+0.03}_{-0.03}$.

\subsection{Relic abundance at tree level}
\label{sec:wo2ndKK}

\begin{figure}[t]
\begin{center}
\scalebox{1.2}{\includegraphics*{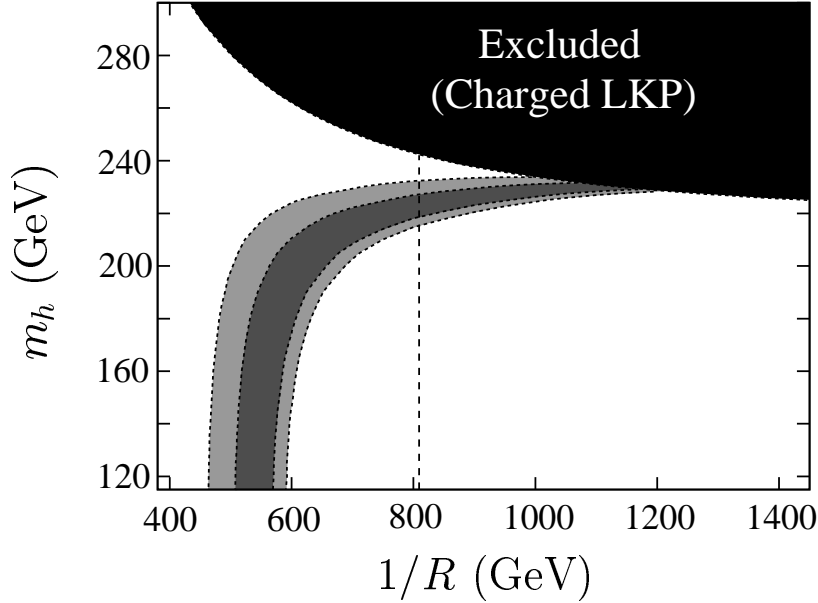}}
\caption{\small The region consistent with the WMAP observation is shown
in $(1/R,m_h)$ plane.
In this calculation, resonance processes are not included.
The dark (light) gray region shows $1 \sigma (2 \sigma )$ allowed region.
In the region $1/R \lesssim 800 $ GeV,
the KK graviton may be the LKP.
}
\label{fig:tree20}
\end{center}
\end{figure}

In Fig.~\ref{fig:tree20},
we show the region consistent with the WMAP observation in $(1/R,m_h)$ plane.
The dark (light) gray region shows $ 1 \sigma (2 \sigma )$ allowed region 
by the WMAP observation \cite{WMAP}.
There is the charged LKP region
in which the KK charged Higgs is the LKP as shown in the previous section.
In the region $1/R \lesssim 800 $ GeV,
the KK graviton may be the LKP as discussed in section \ref{sec:graviton}.
For $m_h \gtrsim 200$ GeV,
the compactification scale as large as $1/R \sim 1$ TeV is allowed.
In this region, coannihilation processes including KK Higgs bosons are efficient,
which has been indicated in Ref.~\cite{Matsumoto:2005uh}.
In the following, we explain this region.

Since the self-coupling of the Higgs is proportional to the square of $m_h$,
the annihilation cross sections of the KK Higgs particles increase with $m_h$.
As a result, $\sigma_{\rm eff} $ is increased for large $m_h$.
Moreover, the mass difference between the LKP and the KK Higgs particle
$A^{(1)}(H^{\pm (1)})$ is smaller for larger $m_h$.
Hence, $A^{(1)}$ and $ H^{\pm (1)}$ are still free from a Boltzmann suppression,
even after other KK particles are decoupled when $m_h$ is large enough.
Since $ g_{\rm eff} $ decreases with decoupling of KK leptons,
the contribution from the annihilation of KK Higgs particles to $ \sigma_{\rm eff} $
becomes large.
Furthermore, the annihilation cross sections of the KK Higgs particles
are much larger than those of KK leptons.
Thus, $\langle \sigma_{\rm eff} v \rangle$ quickly grows with decoupling of KK leptons.

\begin{figure}[t]
\begin{center}
\scalebox{.9}{\includegraphics*{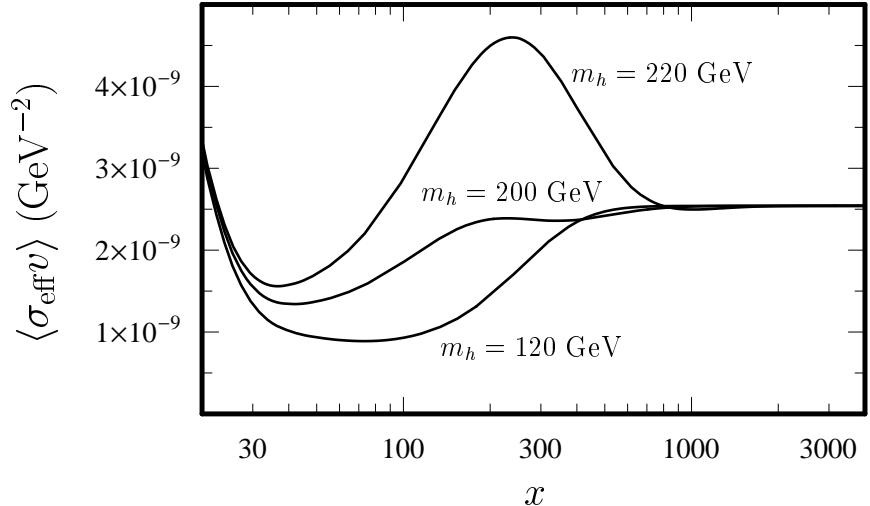}}
\caption{\small The averaged cross sections are shown as a function of $ x $.
The solid lines are for $m_h = 220,200,120$ GeV from top to bottom.
The compactification scale is taken to be $1 / R = 1000$ GeV.
}
\label{fig:avcs}
\end{center}
\end{figure}
\begin{figure}[ht]
\begin{center}
\scalebox{.9}{\includegraphics*{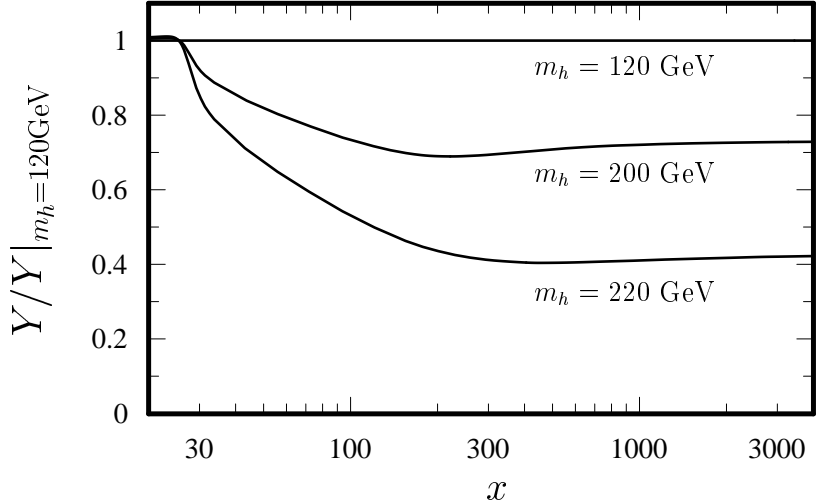}}
\caption{\small The ratio of $Y$ to $Y |_{m_h = 120 \rm GeV} $ are shown
as a function of $x$.
The solid lines are for $m_h = 120,200,220$ GeV from top to bottom.
The compactification scale is taken to be $1 / R = 1000$ GeV.
}
\label{fig:yevo}
\end{center}
\end{figure}

These are illustrated in Figs.~\ref{fig:avcs} and \ref{fig:yevo}.
In Fig.~\ref{fig:avcs}, $ \langle \sigma_{\rm eff} v \rangle $ is plotted
for $ m_h = 220,200,120$ GeV from top to bottom.
The effective annihilation cross sections for $ m_h = 200, 220 $ GeV
increases for $ x = 30 - 200 $ as decouplings of KK leptons,
which continue till $ x \sim 200 $.
Here, the attenuation of $ \langle \sigma_{\rm eff} v \rangle $
for $ m_h =220$ GeV after $x \sim 200 $
is due to the decoupling of $A^{(1)}$ and $ H^{\pm (1)} $.

Due to the rapid growth of the effective annihilation cross section,
sudden freeze-out phenomenon does not occur.
This is a sharp contrast to the case of $m_h =120 $ GeV,
in which sudden freeze-out occurs at $x \sim 25$.
These are shown in Fig.~\ref{fig:yevo}.
In this figure, the ratio, $Y / Y |_{m_h = 120 \rm GeV} $ are shown
as a function of $x$ for $ m_h = 120,200,220$ GeV from top to bottom.
The enhancement of $ \langle \sigma_{\rm eff} v \rangle $
significantly reduces the abundance of dark matter compared to
the case of $m_h = 120 $ GeV.
The late time effect during $ x = 30 - 200$,
reduces the abundance about $20 \% (40 \%)$ for $m_h=200(220)$ GeV
compared to the case of $m_h = 120 $ GeV.

As a result of these effects by the KK Higgs particles,
the relic abundance of the LKP is significantly decreased,
and large $1 / R $ is allowed as shown in Fig.~\ref{fig:tree20}.
We should trace the evolution of the relic abundance of the LKP
until late enough time for the KK Higgs coannihilation region in the MUED model.

\section{Second KK resonances}
\label{sec:2ndKK}

The mass of second KK particle is almost twice mass of its first KK particle.
Thus, in the MUED model, resonance processes
in which the second KK particle propagates in the s-channel are important
for the calculation of the relic abundance.
These effects are partially investigated in Refs.~\cite{Kakizaki:2005en,Kakizaki:2005uy}.
In those works, the second KK particle resonances
are studied for the LKP annihilation and coannihilations
relevant to the SU(2) singlet leptons, $E^{(1)}$.
However, it is found that the second KK resonance processes also play an important role
in coannihilation modes relevant to KK SU(2) doublet leptons 
and KK Higgs particles.

In the following, we discuss second KK resonance processes.
In particular, we focus on the following four processes,
\begin{eqnarray}
\gamma^{(1)} \gamma^{(1)} \to &H^{(2)}& \to \rm SM ~ particles, \\
e^{(1)} \bar e ^{(1)} , \nu^{(1)} \bar \nu ^{(1)} \to &Z^{(2)}&
 \to \rm SM ~ particles, \\
e^{(1)} \bar \nu ^{(1)} \to &W^{-(2)}& \to \rm SM ~ particles, \\
\hspace{-30mm} A^{(1)} A^{(1)}, H^{+ (1)} H^{- (1)} \to &H^{(2)}& \to \rm SM ~ particles.
\label{eq:exception}
\end{eqnarray}

\begin{table}[t]
 \begin{center}
  \caption{\small The annihilation processes in which 
   second KK particles propagate in the s-channel.
   The modes included in our calculation are highlighted with an underbar.
   The processes including $ g^{(1)} , {\rm quark}^{(1)}, W^{(1)} $ and $ Z^{(1)}$
   are omitted due to the reason (a).
   (b): Annihilation is suppressed by small Yukawa coupling;
   (c): Second KK particle decays dominantly into KK particles;
   (d): Second KK particle mass is below the threshold;
   (e): Annihilation is suppressed by the small degree of freedom.
 }
\vspace{3mm}
    \begin{tabular}{|c||c|c|c|c|c|c|c|c|}
     \hline
      & $\gamma^{(1)}$ & $E^{(1)}$ & $e^{(1)}$ & $\nu^{(1)}$
      & $A^{(1)}$ & $H^{\pm(1)}$ & $H^{(1)}$ \\
     \hline \hline
      $\gamma^{(1)}$ & $\underline{H^{(2)}}$ & & & & & & \\
     \hline
      $E^{(1)}$ & \footnotesize $E^{(2)}$(c) & \footnotesize $ \gamma^{(2)}$(d)
      & & & & & \\
     \hline
      $e^{(1)}$ & \footnotesize $e^{(2)}$(c) &
      \begin{tabular}{@{}c@{}}
       \footnotesize $H^{(2)}$(b) \\
       \footnotesize $ A^{(2)}$(b)
      \end{tabular} &
      \begin{tabular}{@{}c@{}}
       $\underline{Z^{(2)}}$ \\
       \footnotesize$ {\gamma}^{(2)}$(d)
      \end{tabular}  & & & & \\
     \hline
      $\nu^{(1)}$ & \footnotesize $\nu^{(2)}$(c) & \footnotesize $ {H}^{\pm(2)}$(b) &
      $\underline{W^{(2)}}$ & 
      \begin{tabular}{@{}c@{}}
       $\underline{Z^{(2)}}$ \\
       \footnotesize$ {\gamma}^{(2)}$(d)
      \end{tabular} & & & \\
     \hline
      $A^{(1)}$ & \footnotesize $H^{(2)} $(e) & \footnotesize $ {e}^{(2)}$(b) &
      \footnotesize $E^{(2)}$(b) & --- & $\underline{H^{(2)}}$ & & \\
     \hline
      $H^{\pm(1)}$ & 
      \begin{tabular}{@{}c@{}}
       \footnotesize $W^{(2)}$(e) \\
       \footnotesize $H^{\pm(2)}$(d)
      \end{tabular}
      & \footnotesize $ {\nu}^{(2)}$(b) & --- & \footnotesize $E^{(2)}$(b) &
      \footnotesize $W^{(2)}$(e)  &
      \begin{tabular}{@{}c@{}}
       \footnotesize $Z^{(2)}$(e) \\
       \footnotesize$ {\gamma}^{(2)}$(d) \\
       $\underline{H^{(2)}}$ 
      \end{tabular} & \\
     \hline
      $H^{(1)}$ & 
      \begin{tabular}{@{}c@{}}
       \footnotesize$Z^{(2)}$(e) \\
       \footnotesize $\gamma^{(2)}$(d) \\
       \footnotesize$A^{(2)}$(d) 
      \end{tabular} &
      \footnotesize ${e}^{(2)}$(b) & \footnotesize ${E}^{(2)}$(b) & --- &
       \begin{tabular}{@{}c@{}}
       \footnotesize ${Z^{(2)}}$(e) \\
       \footnotesize$ {\gamma}^{(2)}$(d) \\
      \footnotesize ${A}^{(2)}$(d)
      \end{tabular} &
      \begin{tabular}{@{}c@{}}
       \footnotesize $W^{(2)}$(e) \\
       \footnotesize $H^{\pm(2)}$(d)
      \end{tabular}
     & \footnotesize$H^{(2)}$(d) \\
     \hline
    \end{tabular}
    \label{table:2ndKK}
  \end{center}
\end{table}

Other second KK processes are not important by the following reasons;
(a) first KK particle in initial state is heavy,
namely processes including $ g^{(1)} , {\rm quark}^{(1)}, W^{(1)}, Z^{(1)}$,
are Boltzmann suppressed;
(b) small Yukawa coupling suppresses processes of
$f^{(1)}+{\rm Higgs}^{(1)} \to f^{\prime (2)} \to $ SM particles ($f$ is a fermion);
(c) second KK particle in the s-channel decays dominantly into
not SM particles but KK particles,
$f^{(1)}+ \gamma^{(1)} \to f^{(2)} \to f^{(1)}+ \gamma^{(1)} $;
(d) the mass of second KK particle is less than
the sum of masses of initial first KK particles;
(e) The annihilation processes including the first KK Higgs 
are suppressed by small degree of freedom.
In the case (e), there are exceptions,
$A^{(1)} A^{(1)}, H^{+ (1)} H^{- (1)} \to H^{(2)} \to $ SM particles processes,
which is discussed later.
In Table~\ref{table:2ndKK},
we summarize the s-channel processes of the second KK particles.

The reasons (a)\footnote{
One might suspect that even if $W^{(1)},Z^{(1)}$ are Boltzmann suppressed,
some resonance processes relevant to these particles enhance $\sigma_{\rm eff}$.
We confirmed numerically these processes are negligibly small.
}, (b), (c) and (d) are obvious.
We should address the reason (e).
At freeze-out, the effective degrees of freedom is about 30, $g_{\rm eff} \sim 30 $,
while $ g_{H^{(1)}}, g_{A^{(1)}},g_{H^{\pm(1)}} = 1$ and $g_{\gamma^{(1)}}=3 $.
Therefore the contribution to $\sigma_{\rm eff}$ from
the annihilation between these particles are suppressed by the factor of $O(10^{-3})$.

However, the reason (e) cannot be applied to the process Eq.~(\ref{eq:exception}).
Because the KK Higgs bosons $A^{(1)}$ and $H^{\pm(1)}$
are highly degenerated with the LKP when $ m_h \gtrsim 200 $ GeV.
After the decoupling of KK leptons,
the KK Higgs particles contribution to $g_{\rm eff}$ is significant,
thus they occupy a half of $g_{\rm eff}$ in the simplest case.
As a result, the enhancement of the KK Higgs annihilation
by the second KK resonances reduces the abundance of dark matter.
Therefore, the second KK resonance for the annihilation of
$A^{(1)} A^{(1)} $ and $H^{+(1)} H^{-(1)} $ should be included.
Here, other resonance processes relevant to $\gamma^{(1)},A^{(1)}$ and $H^{\pm(1)}$,
e.g. $W^{(2)}$ resonance for $ A^{(1)}H^{\pm(1)}$ and $\gamma^{(1)} H^{\pm(1)}$
annihilations, are negligible,
since the collisional energy of annihilating particles is not
sufficient to produce second KK particle after decoupling of KK leptons.

\subsection{$\gamma^{(1)} \gamma^{(1)} \to H^{(2)} \to$ SM particles}

First, we discuss the s-channel $H^{(2)}$ resonance in the LKP annihilation.
Second KK particles cannot decay into the SM particles at tree level,
hence, second KK particle resonance processes are inevitable at loop level.
The interaction induced from radiative correction
is derived by performing loop integrals.
After taking leading parts the following effective interaction turns out to be\footnote{
We correct the coefficient of this interaction in our previous papers
\cite{Kakizaki:2005en,Kakizaki:2005uy}}
\begin{eqnarray}
 {\cal L}_{\rm eff}
 =  \frac{y_t\alpha_s}{3 \pi}
 \ln\left(\Lambda^2R^2\right)
 H^{(2)} \bar{t}t~ .
 \label{eq:h2-SM-SM}
\end{eqnarray}

The decay width of the second KK Higgs particle, 
which is important for calculating the annihilation cross section with the resonance,
is induced by this interaction.
In previous works, we assumed the decay width is given by
\begin{eqnarray}
\Gamma_{H^{(2)}} =
\Gamma^{H^{(2)}}_{t \bar t} = \frac{y_t^2 \alpha_s^2 m_{ H^{(2)} } }{ 24 \pi^3}
\left[ \ln (\Lambda^2 R^2) \right]^2 .
\end{eqnarray}
However, this is not true for $m_h > 2 m_W $,
since $H^{(2)}$ can decay into two KK Higgs particles.
The decay widths for the processes
$H^{(2)} \to H^{+(1)} H^{-(1)}, A^{(1)} A^{(1)} $ are given by
\begin{eqnarray}
\Gamma^{H^{(2)}}_{HH}
 &=& \frac{ \lambda_h^2 m_W^2 }{8 \pi g^2 m_{H^{(2)}}^2 } \sqrt{m_h^2 - 4 m_W^2 }
\label{eq:H2toHpHm},
\\
\Gamma^{H^{(2)}}_{AA}
 &=& \frac{ \lambda_h^2 m_W^2 }{16 \pi g^2 m_{H^{(2)}}^2} \sqrt{m_h^2 - 4 m_Z^2 } .
\label{eq:H2toAA}
\end{eqnarray}
where
$ v^2 = 4 m_W^2 / g^2 $ is used to avoid a confusion between velocity and the vev.
We also used Eqs.~(\ref{eq:higgsmass}), (\ref{eq:chargedhiggsmass})
and (\ref{eq:pseudohiggsmass}).
These decay modes are comparable to that of $t \bar t$ final state
for $m_h > 2 m_W $ even if their phase spaces are suppressed.
Thus, the total decay width is
\begin{eqnarray}
\Gamma_{H^{(2)}}
= \Gamma^{H^{(2)}}_{t \bar t} + \Gamma^{H^{(2)}}_{HH} + \Gamma^{H^{(2)}}_{AA} .
\label{eq:H2all}
\end{eqnarray}

From the above effective interaction in Eq.~(\ref{eq:h2-SM-SM}) 
and the width in Eq.~(\ref{eq:H2all}),
we calculate the annihilation cross section
including the effect of the $H^{(2)}$ resonance \cite{Kakizaki:2005en}:
\begin{eqnarray}
 \sigma^{({\rm Res})}_{\gamma^{(1)} \gamma^{(1)}}
 = \frac{ g^{\prime 4} m_W^2}{18 g^2 \beta m_{H^{(2)}}}
  \frac{\Gamma^{H^{(2)}}_{t \bar t}}
  {(s-m^2_{H^{(2)}})^2 + m_{H^{(2)}}^2 \Gamma_{H^{(2)}}^2}
  \left[ 3 + \frac{s (s - 4 m_{\gamma^{(1)}}^2)}{ 4 m_{\gamma^{(1)}}^4 } \right],
\end{eqnarray}
where
$\beta^2 \equiv \left[ s^2 - 2 (m_1^2 + m_2^2) s + (m_1^2 - m_2^2)^2 \right] / s^2$
for annihilation of particles 1 and 2.
Notice that the interferential contributions between 
tree-level diagrams and one-loop diagrams are negligible
because of the chirality suppression of the top quark mass.

\subsection{ $e^{(1)} \bar e ^{(1)}, \nu^{(1)} \bar \nu ^{(1)} (e^{(1)} \bar \nu ^{(1)})
\to Z^{(2)} (W^{-(2)}) \to$ SM particles }

The KK SU(2) doublet leptons dominate over $ g_{\rm eff}$, therefore,
the resonances in the annihilation processes between these particles are important
for the abundance.

The KK SU(2) gauge boson can decay into KK particles at tree level
and SM particles at loop level as the decay of $H^{(2)}$.
The leading contribution to the decay into the SM particles is dominated by
the process $W^{(2)} \to q \bar q $ mediated by the KK gluon.
Then the effective interaction is 
\begin{eqnarray}
 {\cal L}_{\rm eff}
 = \frac{ 3 \sqrt{2} g \alpha_s}{4 \pi}
 \ln\left(\Lambda^2R^2\right)
 \bar{Q} W^{(2)} {\hspace{-7.5mm} / \hspace{5.2mm}} P_L Q .
 \label{eq:W2-SM-SM}
\end{eqnarray}
Through this interaction,
the decay width of the KK SU(2) gauge boson into SM particles is given by
\begin{eqnarray}
\Gamma^{W^{(2)}}_{\bar Q Q}
= \frac{27 \alpha_2 \alpha_s^2 m_{W^{(2)}} }{32 \pi^2}
 \left[ \ln (\Lambda^2 R^2) \right]^2 ,
\end{eqnarray}
where the decay widths for $W^{(2)}$ and $Z^{(2)}$ are the same
since the weak mixing angle for the KK particle is negligibly small.
Then, the total decay width of the KK SU(2) gauge boson is
\begin{eqnarray}
\Gamma_{W,Z^{(2)}}
&=& \Gamma^{W^{(2)}}_{\bar Q Q} + \Gamma^{W,Z^{(2)}}_{HH,HA}
\nonumber \\
&& + \frac{\alpha_2 m_{W^{(2)}}}{4}
 \left( 1 - \frac{4 m_{l^{(1)}}^2}{m_{W^{(2)}}^2} \right)^{3/2}
\nonumber \\
&& + \frac{\alpha_2 m_{W^{(2)}}}{2}
 \left( 1 - \frac{m_{l^{(2)}}^2}{m_{W^{(2)}}^2} \right)^2 
 \left( 1 + \frac{m_{l^{(2)}}^2}{2 m_{W^{(2)}}^2} \right) ,
\end{eqnarray}
where the third (fourth) term comes from the process $ W^{(2)} \to \bar l ^{(1)} l^{(1)}$
$(W^{(2)} \to \bar l ^{(2)} l^{(0)})$ and
the second term is decay processes into the KK Higgs pair,
which is given by
\begin{eqnarray}
\Gamma^{W^{(2)}}_{HH,HA}
&=& \frac{\alpha_2 m_{W^{(2)}}}{48}
 \left( 1 - 2 \frac{ m_{H^{\pm(1)}}^2 + m_{H^{(1)}}^2}{m_{W^{(2)}}^2} \right)^{3/2} 
\nonumber \\
&& + \frac{\alpha_2 m_{W^{(2)}}}{48}
 \left( 1 - 2 \frac{ m_{H^{\pm(1)}}^2 + m_{A^{(1)}}^2}{m_{W^{(2)}}^2} \right)^{3/2},
\label{eq:WtoHH} \\
\Gamma^{Z^{(2)}}_{HH,HA}
&=& \frac{\alpha_2 m_{W^{(2)}}}{48}
 \left( 1 - 4 \frac{ m_{H^{\pm(1)}}^2 }{m_{W^{(2)}}^2} \right)^{3/2} 
\nonumber \\
&& + \frac{\alpha_2 m_{W^{(2)}}}{48}
 \left( 1 - 2 \frac{ m_{H^{(1)}}^2 + m_{A^{(1)}}^2}{m_{W^{(2)}}^2} \right)^{3/2} ,
\label{eq:ZtoHH}
\end{eqnarray}
where we dropped the square of the small mass difference
$m_{H^{(1)}} -m_{H^{\pm(1)}} $ $(m_{H^{(1)}} - m_{A^{(1)}}) $.
The difference between Eqs.~(\ref{eq:WtoHH}) and (\ref{eq:ZtoHH})
arises from SU(2) breaking effect.

The resonant annihilation cross section is calculated
from these decay widths as,
\begin{eqnarray}
\sigma^{({\rm Res})}_{l^{(1)} \bar l ^{(1)}}
= \frac{ \pi \alpha_2 s \beta }{ 3 m_{W^{(2)}} }
 \frac{\Gamma^{W^{(2)}}_{Q \bar Q}}
 {(s-m^2_{W^{(2)}})^2 + m_{W^{(2)}}^2 \Gamma_{W,Z^{(2)}}^2} ,
\end{eqnarray}
where $\Gamma_{W^{(2)}}(\Gamma_{Z^{(2)}})$
for $e^{(1)} \bar \nu^{(1)} (e^{(1)} \bar e^{(1)} , \nu^{(1)} \bar \nu^{(1)} ) $.
Here, since the interferential contribution between tree and loop diagrams
is subleading, we neglect it.

\subsection{$A^{(1)} A^{(1)}, H^{+ (1)} H^{- (1)} \to H^{(2)} \to$ SM particles}

For $m_h \gtrsim 200$ GeV,
these processes are important as discussed in the beginning of this section. 
The cross sections for the KK charged Higgs pair and for the pseudo KK Higgs pair
are given by the same form as
\begin{eqnarray}
\sigma^{({\rm Res})}_{H^{+(1)} H^{-(1)}}
 = \sigma^{({\rm Res})}_{A^{(1)} A^{(1)}}
 = \frac{ 2 \lambda_h^2 m_W^2 }{g^2 m_{H^{(2)}} \beta }
\frac{\Gamma^{H^{(2)}}_{t \bar t}}{(s- m_{H^{(2)}}^2 )^2 + m_{H^{(2)}}^2 \Gamma_{H^{(2)}}^2}.
\end{eqnarray}
Here, the interferential term between tree and loop diagrams is small
due to the chirality suppression of the top quark.

\subsection{Results}

\begin{figure}[t]
\begin{center}
\scalebox{1.2}{\includegraphics*{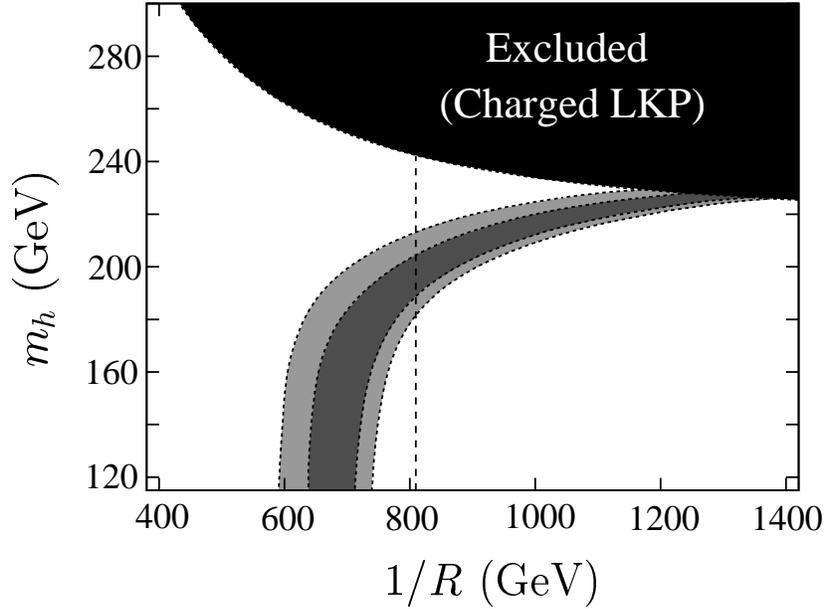}}
\caption{\small The region consistent with the WMAP result is shown
in $(1/R,m_h)$ plane after including the second KK resonances.
The dark (light) gray region shows $1 \sigma (2 \sigma) $ allowed region.
In the region $1/R \lesssim 800 $ GeV,
the KK graviton may be the LKP.
}
\label{fig:res20}
\end{center}
\end{figure}

We show the parameter region of the MUED model consistent with
the WMAP observation in Fig.~\ref{fig:res20}.
The dark (light) gray region shows $1 \sigma (2 \sigma) $ allowed region 
by the observation.
The upper right region is the charged LKP region, which is excluded.
In the region of $1/R \lesssim 800 $ GeV,
the KK graviton may be the LKP, which is discussed in the next section.
By comparing Fig.~\ref{fig:res20} with Fig.~\ref{fig:tree20},
it is found that
the second KK resonances significantly reduce the LKP abundance
and the allowed region of $1 / R $ is increased by about $150-300$ GeV.
In extreme case, $1 /R \sim 1400$ GeV is allowed for $m_h \sim 230$ GeV.
We also calculated the relic abundance in the case $\Lambda R = 50$
and confirmed this result is not changed significantly.

\begin{figure}[t]
\begin{center}
\scalebox{1.}{\includegraphics*{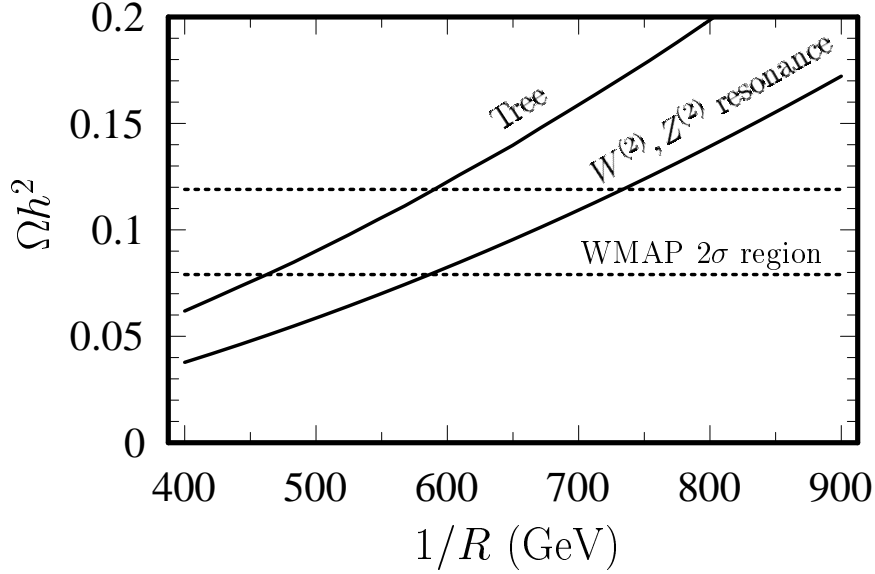}}
\caption{\small The abundance of the LKP as a function of $1/R$
with only the $ W^{(2)}, Z^{(2)}$ resonance and without resonance.
The SM Higgs mass is taken to be $m_h = 120 $ GeV.
The horizontal dotted lines are the allowed region from the WMAP observation
at the $ 2\sigma $ level, $0.079 < \Omega h^2 < 0.119$.
}
\label{fig:2ndKKWZ}
\end{center}
\end{figure}

The shift of the region for $m_h \lesssim 200 $ GeV
originates in the $ W^{(2)}, Z^{(2)}$ resonance.
This is because the KK SU(2) doublet leptons are
dominant components of relic KK particles at freeze-out
and govern $\sigma_{\rm eff}$.
The resonances for the KK SU(2) doublet leptons contribute strongly to
the evolution of the relic abundance of dark matter.
In Fig.~\ref{fig:2ndKKWZ}, the effect of the $ W^{(2)}, Z^{(2)}$ resonance
is illustrated for $m_h = 120$ GeV.
The abundances as a function of $1/R$ are shown as solid lines
for the case with the $ W^{(2)}, Z^{(2)}$ resonances
and that without resonances (tree level result).
The Higgs mass is taken to be $m_h = 120 $ GeV.
Two horizontal lines denote the allowed region from
the WMAP observation at the $ 2\sigma $ level, $0.079 < \Omega h^2 < 0.119$.
The almost all of the shift between Figs.~\ref{fig:tree20} and \ref{fig:res20}
is due to the shift between solid lines in Fig.~\ref{fig:2ndKKWZ}.

\begin{figure}[t]
\begin{center}
\scalebox{1.}{\includegraphics*{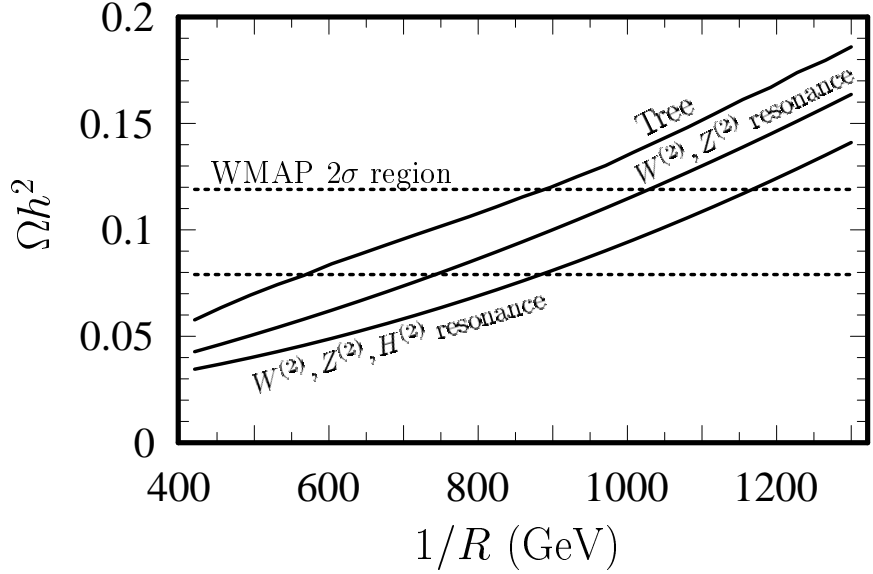}}
\caption{\small The abundance of the LKP as a function of $1/R$
without resonances, with only the $ W^{(2)}, Z^{(2)}$ resonances
and with the $ W^{(2)}, Z^{(2)} , H^{(2)} $ resonances.
The SM Higgs mass is taken to be $m_h = 220 $ GeV.
The horizontal dotted lines are the allowed region from the WMAP observation
at the $ 2\sigma $ level, $0.079 < \Omega h^2 < 0.119$.
}
\label{fig:2ndKKWZH}
\end{center}
\end{figure}

The shift of the region for $m_h \gtrsim 200 $ GeV
is due to the $H^{(2)}$ resonance in addition to $ W^{(2)}, Z^{(2)}$ resonances.
Since the KK Higgs bosons play an important role even at tree level,
the effect of resonances in this coannihilation mode is expected to be important.
In Fig.~\ref{fig:2ndKKWZH}, the effect of the $ W^{(2)}, Z^{(2)}$ resonances
and that of the $ H^{(2)}$ resonance are illustrated.
The abundances as a function of $1/R$ are shown as solid lines
for the case without resonances (tree level result),
with the $ W^{(2)}, Z^{(2)}$ resonances,
and with the $ W^{(2)}, Z^{(2)}$ and $H^{(2)}$ resonances.
The Higgs mass is taken to be $m_h = 220 $ GeV.
Two horizontal lines denote the allowed region from the WMAP observation
at the $ 2\sigma $ level, $0.079 < \Omega h^2 < 0.119$.
About a half of the shift is explained by the $ W^{(2)}, Z^{(2)}$ resonances,
and the rest is due to the $H^{(2)}$ resonance.

\section{KK graviton}
\label{sec:graviton}

We should comment on the KK particle of the graviton.
Since a radiative correction to the mass of the KK graviton
is extremely small due to the gravitational interaction,
the mass is given by $1/R$ with high accuracy.
Therefore, if the mass of the KK photon is larger than $ 1/R $,
the LKP is the KK graviton.
Whether the KK photon mass is larger than the KK graviton mass or not
depends on $1/R$ as seen in Eq.~(\ref{eq:LKP_mass_matrix}).
Since $ \delta m_{W^{(1)}}^2 \gg g' g v^2 / 4 $,
the mass squared difference between the KK photon and graviton
is almost determined by (1,1) component of the mass squared matrix,
$ \delta m_{B^{(1)}}^2 + g^{\prime 2} v^2 / 4 $.
As a result, the LKP can be the KK graviton for $ 1 /R < 809 $ GeV.
It is noted that in the charged LKP region
the KK charged Higgs mass is smaller than $1/R$ and
the LKP is the KK charged Higgs.

The KK graviton LKP leads us to a serious problem.
The lifetime of the Next LKP (NLKP) is very long
due to the gravitational interaction and the small mass difference less than 1 GeV.
The decay of the NLKP produced in the early universe
affects the observation of the cosmic microwave background
or diffuse photon spectrum \cite{Feng:2003xh}.
On the other hand, in the region the KK graviton is not the LKP,
the problem discussed above is replaced with
the problem caused by the late time decay of the KK graviton to the LKP.
This is not serious, since the KK graviton is not abundantly produced by thermal process
if the reheating temperature of the universe is low enough 
\cite{gravitino}.
As a result,
there is the serious problem for the region of $ 1 /R < 809 $ GeV in Fig.~\ref{fig:res20}.

However, there are some setups
in which the KK graviton problem is avoided \cite{Dienes:2001wu,KKgraviton}.
The first setup is based on higher dimensional model than that used in the MUED model, 
and KK particles except the KK graviton are assumed to be localized
in the five dimensional space-time.
In this model, the KK graviton can be heavy without changing other particle spectrum.
Hence, the LKP can be identified with the KK photon
even if $1/R$ is less than 800 GeV.
Another setup is the MUED model with the right-handed neutrino,
which is introduced for the neutrino masses \cite{KKgraviton}.
In UED models, the neutrino mass is given as the Dirac mass,
hence the Yukawa couplings of the right-handed neutrino are extremely small.
In the region that the KK graviton is the LKP,
the NLKP is the KK right-handed neutrino.
Hence, the KK photon decays dominantly into
the KK right-handed neutrino and the ordinary neutrino,
and photons are not emitted in their decays.
Hence, the cosmic microwave background and the diffuse photon spectrum are not distorted.

\section{Summary}
\label{summary}

We have investigated the relic abundance of the LKP dark matter in the MUED model
including the resonance processes by the second KK particles
in all coannihilation processes.
Including the second KK resonance processes reduces the relic abundance,
and the allowed region for $1/R$ consistent with the WMAP observation
shifts by about $150-300$ GeV upward.
In particular, the $W^{(2)}, Z^{(2)}$ resonance processes
in the SU(2) lepton annihilation are responsible for this shift
in the region, $m_h \lesssim 200 $ GeV.
On the other hand, for $m_h \gtrsim 200 $ GeV,
both the $W^{(2)}, Z^{(2)}$ and $H^{(2)}$ resonances contribute comparably.
As a result, the compactification scale consistent with the observed abundance
is $600 {\rm ~GeV} \lesssim 1/R \lesssim 1400 {\rm ~GeV} $ for $\Lambda R = 20$
in the MUED model.

\section*{Acknowledgements}

The work of M.K. was supported in part by
the Japan Society for the Promotion of Science.
The work of S.M. was supported in part by a
Grant-in-Aid of the Ministry of Education,
Culture, Sports, Science, and Technology,
Government of Japan, No. 16081211.


\end{document}